\documentclass[11pt]{amsart}
\usepackage{amssymb,amscd,amsthm,amsmath,amsfonts,epsf,latexsym}
\usepackage[dvips]{graphicx}
\usepackage[all,2cell,ps]{xy}

\input epsf

\theoremstyle{remark}

\theoremstyle{definition}

\numberwithin{equation}{section}

\begin{document}
\title[Topological quantum computing]
{Topologization of electron liquids with Chern-Simons theory and
quantum computation}

\author{Zhenghan Wang}
\email{zhewang@indiana.edu\\zhenghwa@microsoft.com}
\address{Microsoft Project Q\\ c/o Kavli Institute for Theoretical Physics \\
University of California\\ Santa Barbara, CA 93106 \& Department of Mathematics\\
    Indiana University \\
    Bloomington, IN 47405\\
    U.S.A.}

\thanks{The author is partially supported by NSF grant DMS-034772 and EIA 0130388.}

\maketitle

\section{Introduction}
\label{s:intro}

In 1987 a Geometry and Topology year was organized by Prof. Chern
in Nankai and I participated as an undergraduate from the
University of Science and Technology of China.  There I learned
about M. Freedman's work on 4-dimensional manifolds.  Then I went
to the University of California at San Diego to study with M.
Freedman in 1989, and later became his most frequent collaborator.
It is a great pleasure to contribute an article to the memory of
Prof. Chern based partially on some joint works with M. Freedman
and others. Most of the materials are known to experts except some
results about the classification of topological quantum field
theories (TQFTs) in the end. This paper is written during a short
time, so inaccuracies are unavoidable. Comments and questions are
welcome.

There are no better places for me to start than the Chern-Simon
theory. In the hands of Witten, the Chern-Simons functional is
used to define TQFTs which explain the evaluations of the Jones
polynomial of links at certain roots of unity.  It takes great
imagination to relate the Chern-Simons theory to electrons in
magnetic fields, and quantum computing.  Nevertheless, such a
nexus does exist and I will outline this picture.  No attempt has
been made regarding references and completeness.

\section{Chern-Simons theory and TQFTs}

Fix a simply connected Lie group $G$.  Given a closed oriented
3-manifold $M$ and a connection $A$ on a principle $G$-bundle $P$
over $M$, the Chern-Simons 3-form $\textrm{tr}(A\wedge
dA+\frac{2}{3}A^3)$ is discovered when Profs. Chern and Simons
tried to derive a purely combinatorial formula for the first
Pontrjagin number of a 4-manifold.  Let
$CS(A)=\frac{1}{8\pi^2}\int_M \textrm{tr}(A\wedge
dA+\frac{2}{3}A^3)$ be the Chern-Simons functional.  To get a
TQFT, we need to define a complex number for each closed oriented
3-manifold $M$ which is a topological invariant, and a vector
space $V(\Sigma)$ for each closed oriented 2-dimensional surface
$\Sigma$.  For a level $k\geq h^{\vee} +1$, where $h^{\vee}$ is
the dual Coxeter number of $G$, the 3-manifold invariant of $M$ is
the path integral $Z_k(M^3)=\int_{A} e^{2\pi i \cdot k \cdot
CS(A)} DA$, where the integral is over all gauge-classes of
connection on $P$ and the measure $DA$ has yet to be defined
rigorously. A closely related 3-manifold invariant is discovered
rigorously by N. Reshetikhin and V. Turaev based on quantum
groups. To define a vector space for a closed oriented surface
$\Sigma$, let $X$ be an oriented 3-manifold whose boundary is
$\Sigma$.  Consider a principle $G$-bundle $P$ over $X$, fix a
connection $a$ on the restriction of $P$ to $\Sigma$, let
$Z_{k,a}=\int_{(A,a)} e^{2\pi i \cdot k \cdot CS(A)} DA$, where
the integral is over all gauge-classes of connections of $A$ on
$P$ over $X$ whose restriction to $\Sigma$ is $a$. This defines a
functional on all connections $\{a\}$ on the principle $G$-bundle
$P$ over $\Sigma$. By forming formal finite sums, we obtain an
infinite dimensional vector space $S(\Sigma)$. In particular, a
3-manifold $X$ such that $\partial X=\Sigma$ defines a vector in
$S(\Sigma)$. Path integral on disks introduces relations onto the
functionals, we get a finitely dimensional quotient of
$S(\Sigma)$, which is the desired vector space $V(\Sigma)$. Again
such finitely dimensional vector spaces are constructed
mathematically by N. Reshetikhin and V. Turaev. The 3-manifold
invariant of closed oriented 3-manifolds and the vectors spaces
associated to the closed oriented surfaces form part of the
Witten-Reshetikhin-Turaev-Chern-Simons TQFT based on $G$ at
level=$k$. Strictly speaking the 3-manifold invariant is defined
only for framed 3-manifolds.  This subtlety will be ignored in the
following.

Given a TQFT and a closed oriented surface $\Sigma$ with two
connected components $\overline{\Sigma_1}, \Sigma_2$, where
$\overline{\Sigma_1}$ is $\Sigma_1$ with the opposite orientation,
a 3-manifold $X$ with boundary $\partial X=\Sigma$ gives rise to a
linear map from $V(\Sigma_1)$ to $V(\Sigma_2)$.  Then the mapping
cylinder construction for self-diffeomorphisms of surfaces leads
to a projective representation of the mapping class groups of
surfaces. This is the TQFT as axiomatized by M. Atiyah. Later G.
Moore and N. Seiberg, K. Walker and others extended TQFTs to
surfaces with boundaries.  The new ingredient is the introduction
of labels for the boundaries of surfaces.  For the Chern-Simons
TQFTs, the labels are the irreducible representations of the
quantum deformation groups of $G$ at level=$k$ or the positive
energy representations of the loop groups of $G$ at level=$k$. For
more details and references, see [T].

\section{Electrons in a flatland}

Eighteen years before the discovery of electron, a graduate
student E. Hall was studying Electricity and Magnetism using a
book of Maxwell. He was puzzled by a paragraph in Maxwell's book
and performed an experiment to test the statement.  He disproved
the statement by discovering the so-called Hall effect. In 1980,
K. von Klitzing discovered the integer quantum Hall effect (IQHE)
which won him the 1985 Nobel Prize.  Two years later, H. Stormer,
D. Tsui and A. Gossard discovered the fractional quantum Hall
effect (FQHE) which led to the 1998 Nobel Prize for H. Stormer, D.
Tsui and R. Laughlin. They were all studying electrons in a
2-dimensional plane immersed in a perpendicular magnetic field.
Laughlin's prediction of the fractional charge of quasi-particles
in FQHE electron liquids is confirmed by experiments. Such
quasi-particles are anyons, a term introduced by F. Wilczek. Braid
statistics of anyons are deduced, and experiments to confirm braid
statistics are being pursued.

The quantum mechanical problem of an electron in a magnetic field
was solved by L. Landau.  But there are about $10^{11}$ electrons
per $cm^2$ for FQHE liquids, which render the solution of the
realistic Hamiltonian for such electron systems impossible, even
numerically.  The approach in condensed matter physics is to write
down an effective theory which describes the universal properties
of the electron systems. The electrons are strongly interacting
with each other to form an incompressible electron liquid when the
FQHE could be observed. Landau's solution for a single electron in
a magnetic field shows that quantum mechanically an electron
behaves like a harmonic oscillator.  Therefore its energy is
quantized to Landau levels. For a finite size sample of a
2-dimensional electron system in a magnetic field, the number of
electrons in the sample divided by the number of flux quanta in
the perpendicular magnetic field is called the Landau filling
fraction $\nu$. The state of an electron system depends strongly
on the Landau filling fraction. For $\nu< 1/5$, the electron
system is a Wigner crystal: the electrons are pinned at the
vertices of a triangular lattice. For $\nu$ is an integer, the
electron system is an IQHE liquid, where the interaction among
electrons can be neglected. When $\nu$ are certain fractions such
as $1/3, 1/5,...$, the electrons are in a FQHE state. Both IQHE
and FQHE are characterized by the quantization of the Hall
resistance $R_{xy}=\nu^{-1} \frac{h}{e^2}$, where $e$ is the
electron charge and $h$ the Planck constant, and the exponentially
vanishing of the longitudinal resistance $R_{xx}$. There are about
50 such fractions and the quantization of $R_{xy}$ is reproducible
up to $10^{-10}$. How could an electron system with so many
uncontrolled factors such as the disorders, sample shapes and
strength of the magnetic fields, quantize so precisely? The IQHE
has a satisfactory explanation both physically and mathematically.
The mathematical explanation is based on non-commutative Chern
classes.  For the FQHE at filling fractions with odd denominators,
the composite fermion theory based on U(1)-Chern-Simons theory is
a great success: electrons combined with vortices to form
composite fermions and then composite fermions, as new particles,
to form their own integer quantum Hall liquids.  The exceptional
case is the observed FQHE $\nu=5/2$.  There are still very
interesting questions about this FQH state.  For more details and
references see [G].

\section{Topologization of electron liquids}

The discovery of the fractional quantum Hall effect has cast some
doubts on Landau theory for states of matter.  A new concept,
topological order, is proposed by Xiao-gang Wen of MIT.  It is
believed that the electron liquid in a FQHE state is in a
topological state with a Chern-Simons TQFT as an effective theory.
In general topological states of matter have TQFTs as effective
theories. The $\nu=5/2$ FQH electron liquid is still a puzzle. The
leading theory is based on the Pfaffian states proposed by G.
Moore and N. Read in 1991 [MR]. In this theory, the
quarsi-particles are non-abelian anyons (a.k.a. plectons) and the
non-abelian statistics is described by the Chern-Simons-SU(2) TQFT
at level=2.

To describe the new states of matter such as the FQH electron
liquids, we need new concepts and methods.  Consider the following
Gedanken experiment: suppose an electron liquid is confined to a
closed oriented surface $\Sigma$, for example a torus. The lowest
energy states of the system form a Hilbert space $V(\Sigma)$,
called the ground states manifold. In an ordinary quantum system,
the ground state will be unique, so $V(\Sigma)$ is 1-dimensional.
But for topological states of matter, the ground states manifold
is often degenerate (more than 1-dimensional), i.e. there are
several orthogonal ground states with exponentially small energy
differences. This ground states degeneracy is a new quantum
number. Hence a topological quantum system assigns each closed
oriented surface $\Sigma$ a Hilbert space $V(\Sigma)$, which is
exactly the rule for a TQFT. FQH electron liquid always has an
energy gap in the thermodynamic limit which is equivalent to the
incompressibility of the electron liquid. Therefore the ground
states manifold is stable if controlled below the gap. Since the
ground states manifold has the same energy, the Hamiltonian of the
system restricted to the ground states manifold is 0, hence there
will be no continuous evolutions. This agrees with the direct
Lengendre transform form the Chern-Simons Lagrangians to
Hamiltoninans.  Since the Chern-Simons 3-form has only first
derivatives, the corresponding Hamiltonian is identically 0. In
summary, ground states degeneracy, energy gap and the vanishing of
the Hamiltonian are all salient features of topological quantum
systems.

Although the Hamiltonian for a topological system is identically
0, there are still discrete dynamics induced by topological
changes.  In this case the Schrodinger equation is analogous to
the situation for a function $f(x)$ such that $f'(x)=0$, but there
are interesting solutions if the domain of $f(x)$ is not connected
as then $f(x)$ can have different constants on the connected
components. This is exactly why braid group representations arise
as dynamics of topological quantum systems.

\section{Anyons and braid group representations}

Elementary excitations of FQH liquids are quasi-particles.  In the
following we will not distinguish quasi-particles from particles.
 Actually it is not inconceivable that particles are just quasi-particles
 from some complicated vacuum systems.  Particle types serve as
 the labels for TQFTs.
 Suppose a topological quantum system confined on a surface
 $\Sigma$ has elementary excitations localized
 at certain points $p_1,p_2,\cdots$
on $\Sigma$, the ground states of the system outside some small
neighborhoods of $p_i$  form a Hilbert space.  This Hilbert space
is associated to the surface with the small neighborhoods of $p_i$
deleted and each resulting boundary circle is labelled by the
corresponding particle type. Although there are no continuous
evolutions, there are discrete evolutions of the ground states
induced by topological changes such as the mapping class groups of
$\Sigma$ which preserve the boundaries and their labels. An
interesting case is the mapping class groups of the disk with $n$
punctures---the famous braid groups on $n$-strands, $B_n$.

Another way to describe the braid groups $B_n$ is as follows:
given a collection of $n$ particles in the plane ${\mathbb{R}}^2$,
and let $I=[t_0,t_1]$ be a time interval. Then the trajectories of
the particles will be $n$ disjoint curves in ${\mathbb{R}}^2
\times I$ if at any moment the $n$ particles are kept apart from
each other.  If the $n$ particles at time $t_1$ return to their
initial positions at time $t_0$ as a set, then their trajectories
form an $n$-braid $\sigma$. Braids can be stacked on top of each
other to form the braid groups $B_n$. Suppose the particles can be
braided adiabatically so that the quantum system would be always
in the ground states, then we have a unitary transformation from
the ground states at time $t_0$ to the ground states at time
$t_1$. Let $V(\Sigma)$ be the Hilbert space for the ground states
manifold, then a braid induces a unitary transformation on
$V(\Sigma)$.  Actually those unitary transformations give rise to
a projective representation of the braid groups.  If the $n$
particles are of the same type, the resulting representations of
the braid groups will be called the braid statistics.  Note that
there is a group homomorphism from the braid group $B_n$ to the
permutation groups $S_n$ by remembering only the initial and final
positions of the $n$ particles.

The plane ${\mathbb{R}}^2$ above can be replaced by any space $X$
and statistics can be defined for particles in $X$ similarly. The
braid groups are replaced by the fundamental groups $B_n(X)$ of
the configuration spaces $C_n(X)$. If $X={\mathbb{R}}^m$ for some
$m>2$, it is well known that $B_n(X)$ is $S_n$.  Therefore, all
particle statistics will be given by representations of the
permutation groups.  There are two irreducible 1-dimensional
representations of $S_n$, which correspond to bosons and fermions.
If the statistics does not factorize through the permutation
groups $S_n$, the particles are called anyons. If the images are
in $U(1)$, the anyon will be called abelian, and otherwise
non-abelian. The quasi-particles in the FQH liquid at $\nu=1/3$
are abelian anyons.  To be directly useful for topological quantum
computing, we need non-abelian anyons.  Do non-abelian anyons
exist?

Mathematically are there unitary representations of the braid
groups?  There are many representations of the braid groups, but
unitary ones are not easy to find.  The most famous
representations of the braid groups are probably the Burau
representation discovered in 1936, which can be used to define the
Alexander polynomial of links, and the Jones representation
discovered in 1981, which led to the Jones polynomial of links. It
is only in 1984 that the Burau representation was observed to be
unitary by C. Squier, and the Jones representation is unitary as
it was discovered in a unitary world [J1]. So potentially there
could be non-abelian anyon statistics. An interesting question is:
given a family of unitary representations of the braid groups
$\rho_n: B_n\rightarrow U(k_n)$, when this family of
representations can be used to simulate the standard quantum
circuit model efficiently and fault tolerantly?  A sufficient
condition is that they come from a certain TQFT, but is it
necessary?

Are there non-abelian anyons in Nature? This is an important
unknown question at the writing.  Experiments are underway to
confirm the prediction of the existence in certain FQH liquids
[DFN]. Specifically the FQH liquid at $\nu=5/2$ is believed to
have non-abelian anyons whose statistics is described by the Jones
representation at the 4-th root of unity.  More generally N. Read
and E. Rezayi conjectured that the Jones representation of the
braid groups at r-th root of unity describes the non-abelian
statistics for FQH liquids at filling fractions
$\nu=2+\frac{k}{k+2}$, where $k=r-2$ is the level [RR]. For more
details and references on anyons see [Wi].

As an anecdote, a few years ago I wrote an article with others
about quantum computing using non-abelian anyons and submitted it
to the journal Nature.  The paper was rejected within almost a
week with a statement that the editors did not believe in the
existence of non-abelian anyons. Fortunately the final answer has
to come from Mother Nature, rather than the journal Nature.

\section{Topological quantum computing}

In 1980s Yu. Manin and R. Feynman articulated the possibility of
computing machines based on quantum physics to compute much faster
than classical computers.  Shor's factoring algorithm in 1994 has
dramatically changed the field and stirred great interests in
building quantum computers.  There are no theoretical obstacles
for building quantum computers as the accuracy threshold theorem
has shown. But decoherence and errors in implementing unitary
gates have kept most experiments to just a few qubits. In 1997 M.
Freedman proposed the possibility of TQFT computing [F].
Independently A. Kitaev proposed the idea of fault tolerant
quantum computing using anyons [K]. The two ideas are essentially
equivalent as we have alluded before. Leaving aside the issue of
discovering non-abelian anyons, we may ask how to compute using
non-abelian anyons? For more details and references see [NC].

\subsection{Jones representation of the braid groups}

Jones representation of the braid groups is the same as the
Witten-Reshetikhin-Turaev-SU(2) TQFT representation of the braid
groups.  Closely related theories can be defined via the Kaffuman
bracket.  For an even level $k$, the two theories are essentially
the same, but for odd levels the two theories are distinguished by
the Frobenius-Schur indicators.  However the resulting braid group
representations are the same.  Therefore we will describe the
braid group representations using the Kauffman bracket. The
Kauffman bracket is an algebra homomorphism from the group
algebras of the braid groups $\mathbb{C}[B_n]$ to the generic
Temperley-Lieb algebras. For applications to quantum computing we
need unitary theories. So we specialize the Kauffman variable $A$
to certain roots of unity. The resulting algebras are reducible.
Semi-simple quotients can be obtained by imposing the Jones-Wenzl
idempotents. The semi-simple quotient algebras will be called the
Jones algebras, which are direct sum of matrix algebras. Fix $r$
and an $A$ satisfying $A^4=e^{\pm 2\pi i/r}$, the Jones
representation for a braid $\sigma$ is the Kauffman bracket image
in the Jones algebra.  To describe the Jones representation, we
need to find the decomposition of the Jones algebras into their
simple matrix components (irreducible sectors). The set of
particle types for the Chern-Simons-SU(2) TQFT at level=$k$ is
$L=\{0,1,\cdots, k\}$. The fusion rules are given by $a\otimes
b=\oplus c$, where $a,b,c$ satisfy

1). the sum $a+b+c$ is even,

2). $a+b\geq c, b+c\geq a, c+a\geq b$,

3). $a+b+c\leq 2k$.

A triple $(a,b,c), a,b,c\in L$ satisfying the above three conditions
will be called admissible.

The Jones algebra at level=$k$ for n-strands decomposes into
irreducible sectors labeled by an integer $m$ such that $m\in L,
m=n\; mod \; 2$.  Fix $m$, the irreducible sector has a defining
representation $V^m_{1^n}$ with a basis consisting of admissible
labelings of the following tree (Fig. \ref{basis}):

\begin{figure}[tbh]
\includegraphics[width=2.45in]{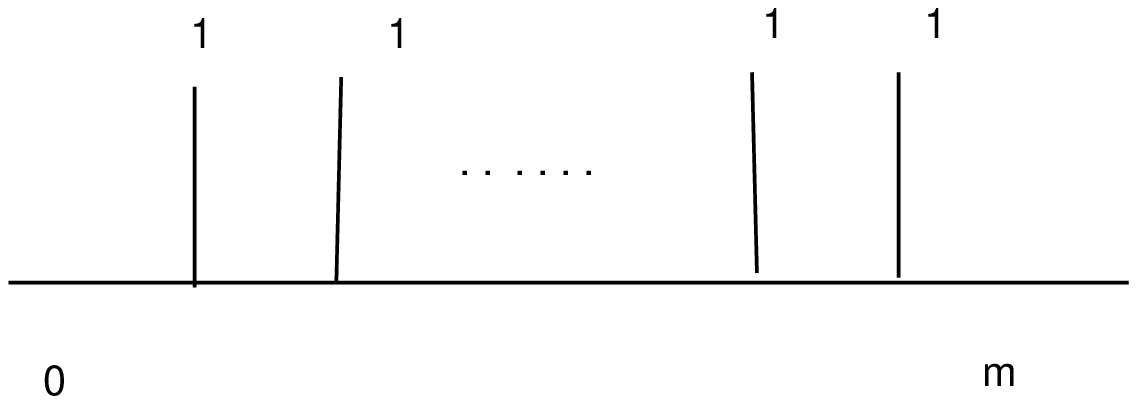}
\caption{Basis}\label{basis}
\end{figure}

There are $n$ vertical edges labeled by 1, and the 0-th horizontal
edge (leftmost) is always labeled by 0, and the n-th edge
(rightmost) is always labeled by $m$.  The internal $(n-1)$ edges
are labeled by $a,b,c,\cdots $ such that any three labels incident
to a trivalent vertex form an admissible triple.  A basis with
internal labelings $a,b,c,\cdots $ will be denoted by
$e^m_{a,b,c,\cdots}$. The Kauffman bracket is $\sigma_i=A\cdot
\textrm{id} +A^{-1}\cdot U_i$, so it suffices to describe the
matrix for $U_i$ with basis $e^m_{a,b,c,\cdots}$ in $V^m_{1^n}$.
The matrix for $U_i$ consists of $1\times 1$ and $2\times 2$
blocks. Fix $m$ and a basis element $e^m_{a,b,c,\cdots}$, suppose
that the $i,i+1,i+2$ internal edges are labeled by $f,g,h$.  If
$f\neq h$, then $U_i$ maps this basis to 0. If $f=h$, then by the
fusions rules $g=f\pm 1$ (the special case is $f=0$, then $g=1$
only), then $U_i$ maps $e^m_{\cdots, f,f\pm1,f,\cdots}$ back to
themselves by the following $2\times 2$ matrix:

$$ \left( \begin{array}{cc} \frac{\Delta_{f+1}}{\Delta_{f}} & x \\
y & \frac{\Delta_{f-1}}{\Delta_f} \end{array} \right),
$$

where $\Delta_k$ is the Chebyshev polynomial defined by
$\Delta_0=1, \Delta_1=d, \Delta_{k+1}=d\Delta_k +\Delta_{k-1},
d=-A^2-A^{-2}$, and $x,y$ satisfy
$xy=\frac{\Delta_{f+1}\Delta_{f-1}}{\Delta_f^2}$.

From those formulas, there is a choice of $x,y$ up to a scalar,
and in order to get a unitary representation, we need to choose
$A$ so that the $2\times 2$ blocks are real symmetric matrices.
This forces $A$ to satisfy $q=A^4=e^{\pm 2\pi i/r}$.  It also
follows that the eigenvalues of $\sigma_i$ are $-1,q$ up to
scalars.

\subsection{Anyonic quantum computers}

We will use the level=2 theory to illustrate the construction of
topological quantum computers. There are three particle types
$\{0,1,2\}$. The label 0 denotes the null-particle type, which is
the vacuum state. Particles of type 1 are believed to be
non-abelian anyons.  Consider the unitary Jones representation of
$B_4$, the irreducible sector with $m=0$ has a basis
$\{e^0_{1,b,1}\}$, where $b=0$ or $2$.  Hence this can be used to
encode a qubit. For $B_6$, a basis consists of
$e^0_{1,b_1,1,b_2,1}$, where $b_i,i=1,2$ is $0$ or $2$.  Hence
this can be used to encode 2-qubits.  In general n-qubits can be
encoded by the $m=0$ irreducible sector of the Jones
representation $\rho^0_{2n+2}$ of $B_{2n+2}$. The unitary matrices
of the Jones representations $\rho_4^0(B_4), \rho_4^0(B_6)$ will
be quantum gates.  To simulate a quantum circuit on n-qubits $U_L:
{(\mathbb{C}^2)}^{\otimes n}\rightarrow {(\mathbb{C}^2)}^{\otimes
n}$, we need a braid $\sigma\in B_{2n+2}$ such that the following
diagram commutes:

\begin{displaymath}
\xymatrix{ ({\mathbb{C})^2}^{\otimes n} \ar[r]^{\cong} \ar[d]_{U_L}
& V^0_{1^{2n+2}} \ar[d]^{\rho^0_{2n+2}(\sigma)}\\
 ({\mathbb{C})^2}^{\otimes n} \ar[r]_{\cong} &  V^0_{1^{2n+2}} }
\end{displaymath}

This is not always possible because the images of the Jones
representation of the braid groups at $r=4$ are finite groups. It
follows that the topological model at $r=4$ is not universal.  To
get a universal computer, we consider other levels of the
Chern-Simons-SU(2) TQFT.  The resulting model for $r=4$ is
slightly different from the above one.  To simulate n-qubits, we
consider the braid group $B_{4n}$. The 4n edges besides the
leftmost in Fig. \ref{basis} can be divided into n groups of 4.
Consider the basis elements such every 4k-th edge is labelled by
0, and every (4k+2)-th edge can be labeled either by 0 or 2. Those
$2^n$ basis elements will be used to encode n-qubits. The
representations of the braid groups $B_{4n}$ will be used to
simulate any quantum circuits on n-qubits.  This is possible for
any level other than 1,2 and 4 [FLW1][FLW2].

\subsection{Measurement in topological models}

A pictorial illustration of a topological quantum computer is as
follows (Fig. \ref{tqc}):

\begin{figure}[tbh]
\includegraphics[width=1.45in]{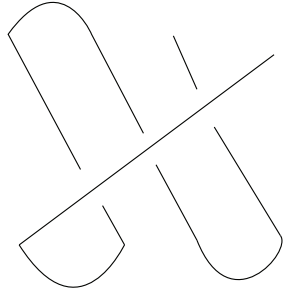}
\caption{Topological model}\label{tqc}
\end{figure}

\begin{figure}[tbh]
\includegraphics[width=3.45in]{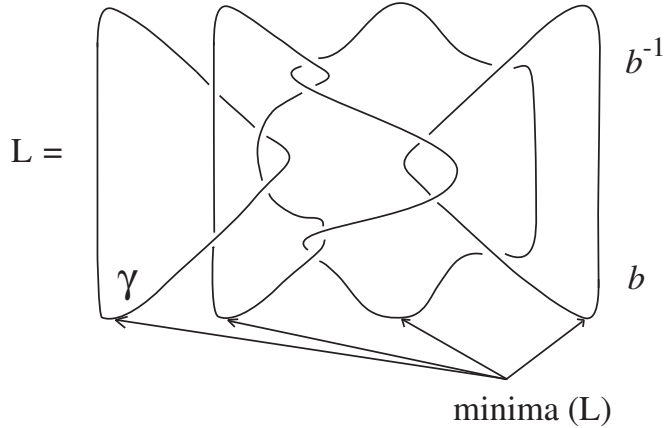}
\caption{Circuit link} \label{link}
\end{figure}

We start the computation with the ground states of a topological
system, then create particle pairs from the ground states to
encode the initial state which is denoted by $|cup>$(two bottom
cup). A braid $b$ is adiabatically performed to induce the desired
unitary matrix $\rho(b)$. In the end, we annihilate the two
leftmost quasi-particles (the top cap) and record the particle
types of the fusion. Then we repeat the process polynomially many
times to get an approximation of the probability of observing any
particle type. Actually we need only to distinguish the trivial
versus all other non-trivial particle types.  For level=3 or
$r=5$, the probability to observe the trivial particle type 0 is
$<cap|\rho^+(b)\prod_0 \rho(b)|cup>$, which is related to the
Jones polynomial of the following circuit link (Fig. \ref{link})
by the formula:

$$p=\textrm{prob}(0)=\frac{1}{1+[2]^2} (1+\frac{(-1)^c\cdot V_L(e^{2\pi
i/5})}{[2]^{c-2}}),$$

where in the formula $c=c(L)$ is the number of components of the
link L, $[2]=-A^2-A^{-2}$ is quantum 2 ar $r=5$.  Our
normalization for the Jones polynomial is that for the unlink with
c components, the Jones polynomial is $(-[2])^c$.

To derive this formula, we assume the writhe of L is 0.  Other
cases are similar.  In the Kauffman bracket formulation, the
projector to null particle type $\prod_0$ is the same as the
element $\frac{U_1}{-[2]}$ of the Jones algebras. It follows that
$p$ is just the Kauffman bracket of the tangle $b\cdot U_1 \cdot
b^{-1}$ divided by $-[2]$. Now consider the Kaffman bracket $<L>$
of $L$, resolving the 4 crossings of $L$ on the component $\gamma$
using the Kauffman bracket results a sum of 16 terms. Simplifying,
we get
$$<L>=(-[2])^c ([2]^2-3)+(4-[2]^2)(-[2])^c\cdot p.$$

Since the writhe is assumed to be 0, the Kauffman bracket is the
same as the Jones polynomial of $L$.  Solving for $p$, we obtain

$$ p=\frac{3-[2]^2}{4-[2]^2}(1+\frac{(-1)^c\cdot V_L(e^{2\pi
i/5})}{[2]^c\cdot (3-[2]^2)}.$$

Direct calculation using the identity $[2]^2=1+[2]$ gives the
desired formula.  This formula shows that if non-abelian anyons
exist to realize the Jones representation of the braid groups,
then quantum computers will approximate the Jones polynomial of
certain links. So the Jones polynomial of links are amplitudes for
certain quantum processes [FKLW].  This inspired a definition of a
new approximation scheme: the additive approximation which might
lead to a new characterization of the computational class BQP
[BFLW].

\subsection{Universality of topological models}

In order to simulate all quantum circuits, it suffices to have the
closed images of the braid groups representations containing the
special unitary groups for each representation space.  In 1981
when Jones discovered his revolutionary unitary representation of
the braid groups, he proved that the images of the irreducible
sectors of his unitary representation are finite if $r=1,2,3,4,6$
for all $n$ and $r=10$ for $n=3$.  For all other cases the closed
images are infinite modulo center.  He asked what are the closed
images? In the joint work with M. Freedman, and M. Larsen [FLW2],
we proved that they are as large as they can be: always contain
the special unitary groups.  As a corollary, we have proved the
universality of the anyonic quantum computers for $r\neq
1,2,3,4,6$.

The proof is interesting in its own right as we formulated a
two-eigenvalue problem and found its solution [FLW2].  The
question of understanding TQFT representations of the mapping
class groups are widely open.  Partial results are obtained in
[LW].

\subsection{Simulation of TQFTs}

In another joint work with M. Freedman, and A. Kitaev [FKW], we
proved that any unitary TQFT can be efficiently simulated by a
quantum computer.  Combined with the universality for certain
TQFTs, we established the equivalence of TQFT computing with
quantum computing.  As corollaries of the simulation theorem, we
obtained quantum algorithms for approximating quantum invariants
such as the Jones polynomial.  Jones polynomial is a
specialization of the Tutte polynomial of graphs.  It is
interesting to ask if there are other partition functions that can
be approximated by quantum computers efficiently [Wel].

\subsection{Fault tolerance of topological models}

Anyonic quantum computers are inherently fault tolerant [K].  This
is essentially a consequence of the disk axiom of TQFTs if the
TQFTs can be localized to lattices on surfaces. Localization of
TQFTs can also be used to establish an energy gap rigorously.

\section{Classification of topological states of matter}

Topological orders of FQH electron liquids are modelled by TQFTs.
It is an interesting and difficult problem to classify all TQFTs,
hence topological orders.  In 2003 I made a conjecture that if the
number of particle types is fixed, then there are only finitely
many TQFTs.   The best approach is based on the concept of modular
tensor category (MTC) [T][BK].  A modular tensor category encodes
the algebraic data inside a TQFT, and describes the consistency of
an anyonic system.  Modular tensor category might be a very useful
concept to study topological quantum systems.  In 2003 I gave a
lecture at the American Institute of Mathematics to an audience of
mostly condensed matter physicists. It was recognized by one of
the participants, Prof. Xiao-gang Wen of MIT, that indeed  tensor
category is useful for physicists as his recent works have shown.

Recently S. Belinschi, R. Stong, E. Rowell and myself have
achieved the classification of all MTCs up to 4 labels.  The
result has not been written up yet, but the list is surprisingly
short.  Each fusion rule is realized by either a Chern-Simons TQFT
and its quantum double.  For example, the fusion rules of
self-dual, singly generated modular tensor categories up to rank=4
are realized by: $SU(2)$ level=1, $SO(3)$ level=3, $SU(2)$
level=2, $SO(3)$ level 5, $SU(2)$ level=3.

\section{Open questions}

There are many open problems in the subject and directions to
pursue for mathematicians, physicists and computer scientists. We
just mention a few here.  The most important for the program is
whether or not there are non-abelian anyons in Nature. Another
question is to understand the boundary (1+1) quantum field
theories of topological quantum systems.  Most of the boundary
QFTs are conformal field theories.  What is the relation of the
boundary QFT with the bulk TQFT?  How do we classify them?

Quantum mechanics has been incorporated into almost every physical
theory in the last century.  Mathematics is experiencing the same
now.  Wavefunctions may well replace the digital numbers as the
new notation to describe our world. The nexus among quantum
topology, quantum physics and quantum computation will lead to a
better understanding of our universe, and Prof. Chern would be
happy to see how important a role that his Chern-Simons theory is
playing in this new endeavor.

\end{document}